\documentstyle[a4]{article}
\title{A new approach to relevancy in Internet searching - the ``Vox Populi Algorithm"}
 \author{
Andreas Schaale ${}^1$, Carsten Wulf-Mathies ${}^2$, S\"onke
Lieberam-Schmidt${}^3$
\\
\\
\small \it
 ${}^1$ Contraco Consulting and Software Ltd., Diepenseer
Str. 10, 15732 Waltersdorf, Germany
\hfill\\
\small \it
 ${}^2$ T-Online International AG, Waldstr. 3, 64331 Weiterstadt, Germany
\hfill\\
\small \it ${}^3$  Universit\"at Siegen, FB 5, H\"olderlinstr. 3,
57068 Siegen, Germany}
\begin{document}
\begin{titlepage}
\maketitle
\begin{abstract}
In this paper we will derive a new algorithm for Internet
searching. The main idea of this algorithm is to extend the
existing algorithms by a component, which reflects the interests
of the users more than existing methods. The ``Vox Populi
Algorithm" (VPA) \cite{patent} creates a feedback from the users
to the content of the search index. The information derived from
the users query analysis is used to modify the existing crawling
algorithms. The VPA controls the distribution of the resources of
the crawler. Finally, we also discuss methods of suppressing
unwanted content (spam). This is necessary in order to enable an
efficient performance of the VPA.
\end{abstract}
\end{titlepage}
The retrieval of relevant information from data sources with a
very complex structure has become a challenging task since the
number of documents in the Internet has reached a level of about
multi billions of documents. Only a small part of them is visible
in search engines. The problem of organizing and structuring these
data into catalogues or searchable databases is of theoretical and
significant practical (commercial) interest.
\\ \\
Let us define the basic components for the mathematical
description of the interests of the users, the relevancy of the
search results and the crawling process. The users of search
engines express their needs for information through the queries
which they address to a searchable database (index) $I$. Each of
the $k$ queries consists of one or more keywords $q$ addressed to
this index. It will be presented as:
\begin{equation}
{\vec{q}_k} = (q_1, ..., q_n)_k
 \label{keyworddef}
\end{equation}
$n$ is the length of the query $k$. The number of keywords per
average query is $n \approx 2$ (status in 2003). The users are
searching for documents $d_j$ (HTML pages, tables, text processing
documents, pictures, multimedia files, ...) containing
information. These documents are grouped (organized) in domains
$D_k$ presenting sets of documents under a common editorial
responsibility and address (URL):
\begin{equation}
D_k = \bigcup^{(n_k)} d_j^{(k)} \;\;\; \mbox{$n_k$ = number of
documents in }D_k
 \label{domaindef}
\end{equation}
The number of domains is about 6.4 million in Germany \cite{denic}
and the number of documents per domain $n_k$ is in the interval
$~10^{0...8}$.
\\ \\
Each document $d$ contains searchable information, today limited
to text information. Content, which is hidden for the {\it
today´s} search technology in non indexable formats (bitmaps,
scripts etc.) will be neglected here and in the following. A
document is characterized by the content of keywords $q$ and the
position of the keyword in certain format elements $e_i$
(metatags, headers, tables, link text etc.):
\begin{equation}
d^{(k)} = f(q_1, q_2, ...,e_1, e_2, ...)
 \label{docdef}
\end{equation}
During the crawling and indexing process, the image of the
document $\hat{d}$ in the searchable index $I$ contains a reduced
set of information - the keywords and their position in the format
elements $e$ of the document. When a query is addressed to the
index $I$ a ranking algorithm generates a set of documents (links)
which is ordered by the relevancy of the found documents. In order
to describe the document ranking process which generates the set
of results on each query, one has to introduce the density $\rho$
of keywords within the documents:
\begin{equation}
\rho^j_i = \frac{n_{q_i}}{n_{e_j}} \label{densitydef}
\end{equation}
where $n_{q_i}$ is the number of the occurrences of the keyword
$q_i$ in the format element $e_j$ and $n_{e_j}$ is the total
number of words in this format element.
\\ \\
Today there exist two basic types of ranking algorithms - the
dynamic and the static ranking algorithms. The dynamic rank of a
document depends on two factors only - the keywords $q$ of the
query and the information content of the documents. Expressed in a
"thumb rule": the higher the keyword density in the document the
higher is the dynamic rank of this document. The relevancy
function $R_d$, defining the dynamic rank of a document, can be
written as:
\begin{equation}
R_d(q_1) \propto  \sum_{k=1}^{N} \mu_k \; \rho^k(q_1)
\;\;\;\mbox{N - number of format elements} \; e
 \label{dynamicrank}
\end{equation}
for a single keyword query. The coefficients $\mu_k$ are free
parameters, defining the importance or weight of each format
element. For example, the occurrence of a keyword in an URL is
usually much more important than in the text itself
$\mu_{URL}>\mu_{text}$. Queries with multiple keywords can be
written as superpositions of single keyword queries:
\begin{equation}
R^{n}_d(q_1,q_2,...,q_n) = R^1(q_1)R^1(q_2)\cdot ... \cdot
R^1(q_n)
 \label{multi-dynamicrank}
\end{equation}
Usually these functions become modified for different purposes,
such as suppression of unwanted information (spam). Other
modifications can take into account the freshness of the document,
the type of the format or other technical parameter.
\\ \\
The practical work on search engines has shown that using only a
document related, dynamical ranking algorithm is insufficient. In
order to also include the importance or the popularity of a domain
(popularity among the webmasters not necessarily among Internet
users), a new type of algorithms was invented - the static ranking
\cite{brin1}. The static rank $R_s$ of a document $d_i$ is related
to the importance of the corresponding domain, where it is
located. The idea of the static rank of a domain $D$ can be
expressed symbolically in the following form:
\begin{equation}
R_s(D) \propto \sum_{j=1}^{N_j} R_s^j
\label{staticrank}
\end{equation}
where the $R_s^j$ is the static rank of the sites linking to the
domain $D$. $N_j$ is the total amount of external links to a
Domain. In \cite{gloeggler} a more detailed definition of the page
rank formula is given:
\begin{equation}
R_s(D) = (1-d) + d \sum_{j=1}^{N_j} R_s^j M_j^{-1}
\label{staticrankdetailed}
\end{equation}
where $d$ is a free parameter (usually in the region $d~0.85$
\cite{gloeggler}) and $M_j$ is the total number of outgoing links
of the referring site. A detailed discussion of the page rank
algorithm used by Google is also found in \cite{kamvar1} and
\cite{kamvar2}.
\\ \\
The resulting rank of a document is a function of the the dynamic
rank (\ref{dynamicrank}) and the static rank (\ref{staticrank}).
There is no unique or even optimal way of constructing this
function. A reasonable way is to choose the resulting relevancy
$R_{ds}$ as a product of the dynamic and static rank:
\begin{equation}
R_{ds} = R_d(q) \cdot R_s(d_i)  \label{ds-rank}
\end{equation}
Analyzing (\ref{ds-rank}) a usual approach would be using
$R_s(D_i)$ instead of $R_s(d_i)$. In practice the static rank of a
document depends not only on the static rank of the domain $D$
containing $d_i$, but also on the position in the domain (link
topology of the domain). At present this kind of search algorithms
is in use in every major internet search engine.
\\ \\
The algorithms described above do indeed meet the needs of the
users. This approach is reasonable from an academic point of view
and it has produced remarkable results in the past. Today it has
become more difficult to make use of the link topology - very
often the links are not set according to the content relevancy,
but for other (economic) reasons. To the extent that the search
engines have become the most important information retrieval tool,
they have also become a target of spamming (site owners try to
fake the search engines, virtually presenting more important
content than there really is). An effective method of detecting a
certain type of spam is described in the appendix. Applying filter
mechanisms and modifying the parameters of the dynamic and the
static relevancy algorithms, one can ``fine tune" the quality of
the Internet search engines.
\\ \\
The two methods described above explicitly do not take into
account the most important factor, the interest of the users
searching for information. The dynamic and the static relevancy of
a document are influenced by the content of the site and by the
``citation" by other sites. There is no methodical component, that
reflects the voice of the searching people. This will be done by
the ``Vox Populi Algorithm" (people`s voice).
\\ \\
The main idea of the VPA is to use the information that is
extractable from the user query analysis to enhance the quality of
the search. This can be done in two different ways, by modifying
either the ranking or the crawling algorithm. In this paper the
focus is not on the ranking, but on the crawling algorithm. The
crawling algorithm defines which domain and how much of the
content will be included into the search index. Sites which are
not included cannot be found by the best ranking algorithm. At
present there is only a small fraction ($< 10\%$) of the Internet
sites indexed by the search engines. The much bigger part of the
Internet (``Deep Web") is not visible in any of the search
engines.
\\ \\
The source of information is the analysis of the queries
$\vec{q}$, reflecting the users interests and needs. The query set
$Q$ may contain all single and multiple keyword queries of the
users (\ref{keyworddef}). Based on these queries a
multidimensional tensor $\Omega$ can be defined, containing the
information of the multiple keyword correlations with the
dimension $N_{max}$.
\begin{equation}
dim[\Omega(Q)] = N_{max}
 \label{Otensor}
\end{equation}
$N_{max}$ is the maximum length of a query - theoretically it can
be infinite. Practically the amount of queries having $>6$
keywords is $<1\%$, while the average query consists of about
$N=2$ keywords. In order to simplify the further calculations one
can reduce the dimension of (\ref{Otensor}) in the following way:
\begin{equation}
\Omega^{N_{max}}(Q) \rightarrow \Omega^{N=2}(Q) \equiv \Omega
 \label{reducedOtensor}
\end{equation}
In this reduction algorithm, the queries with more than two
keywords are replaced by two keyword queries, containing all
possible paired combinations. For example, a three keyword query
is equivalent to 3 two keyword queries and so on.
\\ \\
The matrix $\Omega$ is a correlation matrix of all keywords of the
query set $Q$, which is analyzed. $\Omega$ is a positive and
symmetric matrix \footnote{The analysis of the order of the
keywords shows a statistical asymmetry for the order of keywords
$N(1,2) \neq N(2,1)$. Users interested in the explicit order of
the keywords can use the option called ``Exact Phrase", which is
available on any modern search engine. Therefore it is reasonable
to assume that the order of the keywords is not important for the
users when they make simple queries (more than 90\% of all queries
are of this type). We will use here the approximation $(1.2) =
(2.1)$}. One can calculate the eigenvectors and eigenvalues of
$\Omega$, transforming it into the diagonal form:
\begin{equation}
K^{-1} \Omega K = \Omega^{diag}
 \label{Otensordiag}
\end{equation}
The details of the diagonalization procedure are well known, see
\cite{bronstein} or any other standard textbook on mathematics. It
is now important to understand the practical meaning of the
matrices $K$ and $\Omega^{diag}$. The matrix $K$ consists of
eigenvectors which are keyword combinations:
\begin{equation}
K=\left(
\begin{array}{c}
\vec{e}^1 \\
\vec{e}^2 \\
\vec{e}^3 \\
...
\end{array}
\right) \label{Kmatrix}
\end{equation}
where each eigenvector has the coordinates
\begin{equation}
\vec{e}^j=(c_1q_1,c_2q_2, ...)^j
 \label{qvector}
\end{equation}
similar to the definition (\ref{keyworddef}) the $q_i$ are the
keywords and the coefficients $c_i^j$ are positive numbers, giving
each keyword some "weight" compared to the other ones (How
frequent do the users ask for this keyword?). The coefficients
determine the relative importance of a keyword within an
eigenvector. A typical eigenvector (or better ``eigenquery") has
the form (based on the data \cite{keyworddatenbank}, Aug. 2003).
\begin{equation}
\vec{e}^j=(``mp3",\;0.73\cdot ``downloads", \; 0.43 \cdot ``free",
...)
 \label{qvectordemo}
\end{equation}
This query shows how the {\it average} user is asking, when he is
searching for mp3 downloads at no cost. The reduced $(N=3)$
keyword matrix of the example above has the form
\cite{keyworddatenbank}:
\begin{equation}
\Omega=
\begin{array}{lccc}
          & mp3       & download  & free     \\
mp3       & 37.2 \%   & 8.8 \%    & 2.7\%    \\
download  & 8.8 \%    & 19.2 \%   & 3.6\%    \\
free      & 2.7 \%    & 3.6 \%    & 13.4\%
\end{array}
\label{mp3matrix}
\end{equation}
The difference between the typical keyword search at present and
our approach is that the words here have different weights,
determining their relative importance for the users.
\\ \\
Another important information about the significance of keyword
combinations is contained in the matrix $\Omega^{diag}$.
\begin{equation}
\Omega^{diag}=\left(
\begin{array}{ccccc}
\lambda_1 & 0         & 0   & ...           & 0 \\
0         & \lambda_2 & 0   & ...           & 0  \\
0         & 0         & ... &  ...          & ...  \\
0         & 0         & ... & \lambda_{N-1} & 0 \\
 0        & 0         & ... & 0             & \lambda_N
\end{array}
\right)
 \label{Omegadiag}
\end{equation}
Each eigenvalue $\lambda_i$ corresponds to an eigenvector in
(\ref{qvector}). The eigenvalue can be interpreted as the
importance of the corresponding eigenvector - it defines the
importance of an eigenquery for the users.
\\ \\
Finally, we have developed the tools for defining how a search
engine can use the information of the users to determine, which
content should be enhanced or reduced in the index. Based on the
described algorithm it is possible to define which content is the
``most wanted" content and which sites deliver this type of
content:
$$
(c_1q_1 + c_2q_2 + ...) \rightarrow \;\;\mbox{search engine}
\rightarrow \;\;\mbox{list of ranked domains}
$$
Crawling the Internet, each domain is given certain resources by
the search engine, such as CPU time and memory in the index
(alternatively also the number of crawled documents or other
parameters, depending on the settings of the search engine).
\\ \\
The practical realization of the VPA as an extension of an
existing Internet search could be performed using the following
procedure:
\begin{enumerate}
\item Generate a ranking of domains, addressing the eigenqueries
(\ref{qvector}) to the existing (old) search index, the priority
of those domains is defined by the size of eigenvalues.
(\ref{Omegadiag}).
 \item Modify the existing resource ranking
list with respect to these eigenvalues.
 \item Use the new determined ranking of the domains for crawling the Internet
 according to the modified resource
 distribution.
 \item Repeat the cycle.
\end{enumerate}
In order to determine which sites best fit the eigenqueries, it is
useful to calculate a dynamic rank for a whole domain, not just
for a single document. A simple method would be to summarize the
total score of all documents in one domain:
\begin{equation}
{\bf R_D}(e_i) \propto  \sum_{k=1}^{N_D} R_d^k(e_i)
 \label{domainrank}
\end{equation}
Let us assume, that the amount of resources (CPU time, number of
documents, data volume etc.) given to each domain, when crawling
it, can be expressed in a function $M$, with
\begin{equation}
M=M(D_k,R_s, ...)
 \label{M1}
\end{equation}
In order to apply the VPA one can modify (\ref{M1}) in the
following way:
\begin{equation}
M\rightarrow \hat{M}=M\cdot R_{VPA}
 \label{M2}
\end{equation}
The function $R_{VPA}$ defines the VPA correction with regard to
the old crawling algorithm. The function $R_{VPA}$ can be
presented in different ways. The basic requirement for the
function is that it is monotone concerning the parameters
$\lambda_i$, which define quantitatively how relevant a query is
for the users. Following Occam`s principle of simplicity
(Pluralitas non est ponenda sine neccesitate - Entities should not
be multiplied unnecessarily) this function should use only a
minimum set of free parameters, which will allow the adoption (or
``fine tuning") the algorithm to the local requirements:
\begin{equation}
R_{VPA}(D_k)=\big( 1+ \alpha \cdot \lambda_k^\beta \big) \;\;\;
\alpha,\beta >0
 \label{RVPA}
\end{equation}
The parameter $\alpha$ and $\beta$ can be chosen freely. In the
limit, the new algorithm generates the existing results in
(\ref{M2}).
\begin{equation}
\lim_{\lambda \rightarrow 0}\hat{M} = M
 \label{Mlimit}
\end{equation}
In this paper we have shown how the analysis of queries can be
used to enhance the relevant and ``most wanted" content in a
search index. In this way the relevancy, experienced by the users
of the search should grow - the users will find more of what they
are interested in. The existing system of the relevancy ranking of
documents or domains can remain unchanged. The algorithm will not
replace existing crawling and ranking algorithms, but the VPA will
extend them by a qualitatively new component.
\newpage
{\bf Appendix}
\\ \\
The static rank algorithm has also become the target of spamming
(for example, ``Google bombing" \cite{googlebombing}). This means
that webmasters are creating clusters of domains, which consist of
very similar sites, referring to a single domain or a document.
This kind of spam cluster can consist of many domains, which do
not contain any valuable content at all. Because of this the
static rank consequently is becoming more and more a measure of
the marketing budget or the cleverness of the webmaster of a
domain, rather than a measure of ``real" reputation or content
quality. As a result of this development, the importance of the
static rank as a tool for determining the quality or the relevancy
of a site is decreasing.
\\ \\
We want to propose an algorithm which identifies this kind of
spamming. The basic idea of the static rank is reasonable - the
more important sites refer (link) to a site, the more important is
the site. There is a way to discriminate between ``natural grown"
link clusters and ``artificial" ones (spam).
\\ \\
In order to find a quantitative method which can discriminate
between these two types of link clusters, one can introduce the
function which describes the statistical distribution of the
relevancy $R^j_s$ of the links, pointing to the document $d_i$:
\begin{equation}
\phi(R^j_s) = e^{-\frac{(R^j_s-R_0)^2}{\sigma^2}}
\label{spamcluster}
\end{equation}
here $R_0$ is the average static rank of all sites, linking to the
center of this cluster $d_i$. The parameter $\sigma$ defines the
width of the distribution.
\\ \\
The above mentioned types of clusters can be discriminated using
the distribution $\phi$ - natural grown clusters contain links
from an inhomogeneous set of sites, for example, the links to a
site of a well known university will come from very small
(amateur) sites of students, employees and alumnies (with a low
page rank), via semi professional institutional sites (spin offs,
research partners, ...) up to sites of other high ranked
universities or institutes. The artificial link cluster consists
of automatically generated sites, each of them usually optimized
for different keywords, but having approximately the same static
rank. As a result of this it is possible to introduce a ``cut off"
criteria based on formula (\ref{spamcluster}). A cluster is most
likely spam, if the condition
\begin{equation}
\sigma_{spam} < \sigma_{critical}
\label{spamcut}
\end{equation}
is fulfilled. Here $\sigma_{critical}$ is an empirical parameter,
which can be determined from the analysis of known natural and
artificial clusters (or from the software generating the sites of
the spam cluster). Estimates have shown that one can expect a
result like $\sigma_{natural} >> \sigma_{artificial}$. A short
test example can demonstrate this: the distribution of the page
ranks of sites linking to the homepage of Steven Hawking
\cite{hawking internet} analyzed based on formula
(\ref{spamcluster}) have a width of $\sigma^2 = 1.1$ , while the
sites belonging to a typical spam cluster have a page rank
distribution with $\sigma^2 = 0.5 ... 0.7 $ \footnote{The data of
this example are based on the page rank indicator of Google
\cite{googlebar}.}. The parameter $\sigma$ can be used for
separating between these two type of link clusters. The data of
this example are based on the indications of the page rank
indicator of Google`s toolbar \cite{googlebar}.

\end{document}